\documentclass[aps,twocolumn, prl]{revtex4}
\usepackage{graphicx}
\usepackage[scanall]{psfrag}
\usepackage{amssymb}

\begin{document}

%
%
%
%
\def\oti{{\otimes}}
\def\lb{ \left[ }
\def\rb{ \right]  }
\def\tilde{\widetilde}
\def\bar{\overline}
\def\hat{\widehat}
\def\*{\star}
\def\[{\left[}
\def\]{\right]}
\def\({\left(}		\def\BL{\Bigr(}
\def\){\right)}		\def\BR{\Bigr)}
	\def\BBL{\lb}
	\def\BBR{\rb}
%
%
\def\zb{{\bar{z} }}
\def\zbar{{\bar{z} }}
\def\frac#1#2{{#1 \over #2}}
\def\inv#1{{1 \over #1}}
\def\half{{1 \over 2}}
\def\d{\partial}
\def\der#1{{\partial \over \partial #1}}
\def\dd#1#2{{\partial #1 \over \partial #2}}
\def\vev#1{\langle #1 \rangle}
\def\ket#1{ | #1 \rangle}
\def\rvac{\hbox{$\vert 0\rangle$}}
\def\lvac{\hbox{$\langle 0 \vert $}}
\def\2pi{\hbox{$2\pi i$}}
\def\e#1{{\rm e}^{^{\textstyle #1}}}
\def\grad#1{\,\nabla\!_{{#1}}\,}
\def\dsl{\raise.15ex\hbox{/}\kern-.57em\partial}
\def\Dsl{\,\raise.15ex\hbox{/}\mkern-.13.5mu D}
%
%
\def\ga{\gamma}		\def\Ga{\Gamma}
\def\be{\beta}
\def\al{\alpha}
\def\ep{\epsilon}
\def\vep{\varepsilon}
\def\la{\lambda}	\def\La{\Lambda}
\def\de{\delta}		\def\De{\Delta}
\def\om{\omega}		\def\Om{\Omega}
\def\sig{\sigma}	\def\Sig{\Sigma}
\def\vphi{\varphi}

%
%
\def\CA{{\cal A}}	\def\CB{{\cal B}}	\def\CC{{\cal C}}
\def\CD{{\cal D}}	\def\CE{{\cal E}}	\def\CF{{\cal F}}
\def\CG{{\cal G}}	\def\CH{{\cal H}}	\def\CI{{\cal J}}
\def\CJ{{\cal J}}	\def\CK{{\cal K}}	\def\CL{{\cal L}}
\def\CM{{\cal M}}	\def\CN{{\cal N}}	\def\CO{{\cal O}}
\def\CP{{\cal P}}	\def\CQ{{\cal Q}}	\def\CR{{\cal R}}
\def\CS{{\cal S}}	\def\CT{{\cal T}}	\def\CU{{\cal U}}
\def\CV{{\cal V}}	\def\CW{{\cal W}}	\def\CX{{\cal X}}
\def\CY{{\cal Y}}	\def\CZ{{\cal Z}}

\def\rvac{\hbox{$\vert 0\rangle$}}
\def\lvac{\hbox{$\langle 0 \vert $}}
\def\comm#1#2{ \BBL\ #1\ ,\ #2 \BBR }
\def\2pi{\hbox{$2\pi i$}}
\def\e#1{{\rm e}^{^{\textstyle #1}}}
\def\grad#1{\,\nabla\!_{{#1}}\,}
\def\dsl{\raise.15ex\hbox{/}\kern-.57em\partial}
\def\Dsl{\,\raise.15ex\hbox{/}\mkern-.13.5mu D}
%
%
%
\font\numbers=cmss12
\font\upright=cmu10 scaled\magstep1
\def\stroke{\vrule height8pt width0.4pt depth-0.1pt}
\def\topfleck{\vrule height8pt width0.5pt depth-5.9pt}
\def\botfleck{\vrule height2pt width0.5pt depth0.1pt}
\def\Zmath{\vcenter{\hbox{\numbers\rlap{\rlap{Z}\kern
0.8pt\topfleck}\kern 2.2pt
                   \rlap Z\kern 6pt\botfleck\kern 1pt}}}
\def\Qmath{\vcenter{\hbox{\upright\rlap{\rlap{Q}\kern
                   3.8pt\stroke}\phantom{Q}}}}
\def\Nmath{\vcenter{\hbox{\upright\rlap{I}\kern 1.7pt N}}}
\def\Cmath{\vcenter{\hbox{\upright\rlap{\rlap{C}\kern
                   3.8pt\stroke}\phantom{C}}}}
\def\Rmath{\vcenter{\hbox{\upright\rlap{I}\kern 1.7pt R}}}
\def\Z{\ifmmode\Zmath\else$\Zmath$\fi}
\def\Q{\ifmmode\Qmath\else$\Qmath$\fi}
\def\N{\ifmmode\Nmath\else$\Nmath$\fi}
\def\C{\ifmmode\Cmath\else$\Cmath$\fi}
\def\R{\ifmmode\Rmath\else$\Rmath$\fi}

\def\barray{\begin{eqnarray}}
\def\earray{\end{eqnarray}}
\def\beq{\begin{equation}}
\def\eeq{\end{equation}}

\def\n{\noindent}

\def\Tr{\rm Tr} 
\def\xvec{{\bf x}}
\def\kvec{{\bf k}}
\def\kvecp{{\bf k'}}
\def\omk{\om{\kvec}} 
\def\dk#1{\frac{d\kvec_{#1}}{(2\pi)^d}}
\def\2pid{(2\pi)^d}
\def\ket#1{|#1 \rangle}
\def\bra#1{\langle #1 |}
\def\vol{V}
\def\adag{a^\dagger}
\def\rme{{\rm e}}
\def\Im{{\rm Im}}
\def\pvec{{\bf p}}
\def\fermiS{\CS_F}
\def\cdag{c^\dagger}
\def\adag{a^\dagger}
\def\bdag{b^\dagger}
\def\vvec{{\bf v}}
\def\muhat{{\hat{\mu}}}
\def\vac{|0\rangle}
\def\pcut{{\Lambda_c}}
\def\chidot{\dot{\chi}}
\def\gradvec{\vec{\nabla}}
\def\psitilde{\tilde{\Psi}}
\def\psibar{\bar{\psi}}
\def\psidag{\psi^\dagger} 
\def\m{m_*}
\def\up{\uparrow}
\def\down{\downarrow}
\def\Qo{Q^{0}}
\def\vbar{\bar{v}}
\def\ubar{\bar{u}}
\def\smallhalf{{\textstyle \inv{2}}}
\def\smallsqrt{{\textstyle \inv{\sqrt{2}}}}
\def\rvec{{\bf r}}
\def\avec{{\bf a}}
\def\pivec{{\vec{\pi}}}
\def\svec{\vec{s}} 
\def\phivec{\vec{\phi}}
\def\daggerc{{\dagger_c}}
\def\Gfour{G^{(4)}}
\def\dim#1{\lbrack\!\lbrack #1 \rbrack\! \rbrack }
\def\qhat{{\hat{q}}}
\def\ghat{{\hat{g}}}
\def\nvec{{\vec{n}}}
\def\bull{$\bullet$}
\def\ghato{{\hat{g}_0}}
\def\r{r}
\def\deltaq{\delta_q}
\def\gcharge{g_q}
\def\gspin{g_s}
\def\deltas{\delta_s}
\def\gQC{g_{AF}} 
\def\ghatqc{\ghat_{AF}}
\def\xqc{x_{AF}}
\def\mhat{\hat{m}}
\def\xup{x_2}
\def\xdown{x_1}
\def\sigmavec{\vec{\sigma}}
\def\xopt{x_{\rm opt}}
\def\Lambdac{{\Lambda_c}}
\def\angstrom{{{\scriptstyle \circ} \atop A}     }
\def\AA{\leavevmode\setbox0=\hbox{h}\dimen0=\ht0 \advance\dimen0 by-1ex\rlap{
\raise.67\dimen0\hbox{\char'27}}A}
\def\ratio{\gamma}

\title{A unique non-Landau/Fermi liquid in $2d$ for high $T_c$
superconductivity}
\author{Eliot Kapit and Andr\'e  LeClair}
\affiliation{Newman Laboratory, Cornell University, Ithaca, NY} 
\date{May 2008}

\bigskip\bigskip\bigskip\bigskip

\begin{abstract}

It is shown that  the main  features of the
high $T_c$ phase diagram  can be calculated as a function
of doping in a simple,  essentially unique non-Landau/Fermi liquid in $2d$ 
with quartic interactions. This depends on a single
parameter $0<\gamma <1$ which encodes the strength of the 
interaction at short distances.        A new d-wave gap equation has
solutions that fall under 
a superconducting dome,  which terminates at the renormalization
group fixed point.   Optimal doping is estimated to  occur just
below  $3/2\pi^2$.   The scale for $T_c$ is set by  the  recently measured universal nodal  Fermi
velocity and lattice spacing,  and is estimated to be
$120K < T_c <  160K$ for LaSrCuO.

\end{abstract}

\maketitle

\begin{figure}[htb] 
\begin{center}
\hspace{-15mm}
\psfrag{X}{$h = {\rm doping}$ }
\psfrag{A}{$0.5$ }
\psfrag{B}{$.06$}
\psfrag{C}{$h_{AF}$}
\psfrag{D}{$h_*$}
\psfrag{F}{$SC$} 
\psfrag{G}{${\rm pseudogap}$}
\psfrag{H}{$T_{pg}$ }
\psfrag{I}{$AF$}
\psfrag{N}{$T_N$}
\psfrag{S}{$T_c$}
\psfrag{T}{$\deltas' ,\deltaq'$}
\psfrag{n}{$h_1$}
\includegraphics[width=6cm]{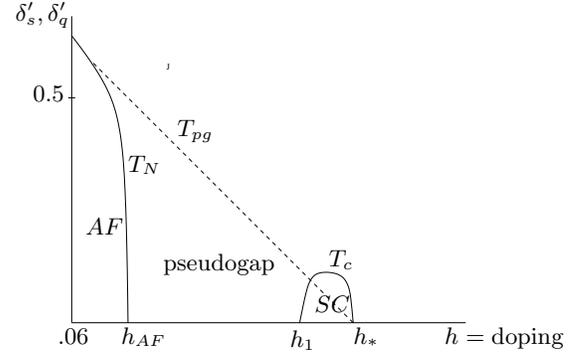} 
\end{center}
\caption{Calculated phase diagram as a function of 
hole doping based on a single parameter $0<\gamma<1$,
set equal to $1$ (infinitely strong coupling at
short distances).  
What is shown are solutions  $\deltas' ,  \deltaq'$   
of the AF and d-wave gap equations below, which are proportional
to the critical temperature.  
$T_{pg}$ is simply the RG scale.   
The AF transition point at $h_{AF} = \frac{3}{4\pi^2} $ 
is first order.   The SC transition at $h_* = \frac{3}{2\pi^2}$
 is second-order
and corresponds to the fixed point of the renormalization group.
$h_1 = h(x_1) \approx 0.13$ is not universal.} 
\vspace{-2mm}
\label{Figure1} 
\end{figure}

There remain many fundamental unanswered questions in the
theory of high $T_c$ superconductivity in the cuprates.     
How can one theory interpolate between anti-ferromagnetic (AF) 
and superconducting (SC) order,  since AF requires repulsive
interactions and SC attractive?   What is the precise mechanism
that gives d-wave pairing?   What is the nature of the pseudogap?  
The most important guide in tackling these problems is the
non-Fermi liquid behavior,  as emphasized early on by Anderson\cite{Anderson}.
The reason is that if one insists on a local quantum field theory 
description in the rotationally invariant  long wavelength limit,  
relevant interactions which lead to a low energy fixed point are
exceedingly constrained and in fact rare.    In this paper we 
describe an essentially unique non-Fermi liquid theory in $2d$
based on 4-fermion interactions which automatically has
$SO(5)$ symmetry.    It thus gives a microscopic model where the
ideas of Zhang may be explored\cite{Zhang}.   
 The requirements of a consistent local  effective field 
theory for expansion around the Fermi surface and relevance of
quartic interactions make this theory inevitable. 
This model was proposed earlier
in some preliminary work by one of us\cite{LeClair1,Neubert} which focused on 
AF order;  at the time the d-wave SC properties were not understood.  
The model is in line with ideas that emphasize the r\^ole of the Hubbard
and Heisenberg models\cite{Anderson2}.  
However our model cannot be simply derived by a naive scaling
limit of the Hubbard model since the latter only has 
at most $SO(4)$ symmetry\cite{YangZhang}.    
Once  the   theory is formulated,
the main properties of high $T_c$  follow  naturally in an  analysis
involving only  simple 1-loop calculations,  which are accurate since 
at the fixed point the coupling is small $\approx \inv{8}$. 
    Our model  reveals that  high $T_c$ SC
may be a  remarkably universal phenomenon that manages to realize some
subtle theoretical loopholes,    and its main properties follow from the existence
of the low energy renormalization group (RG)  fixed point.  
Such quantum critical points were emphasized
by Vojta and Sachdev\cite{Vojta}.    
We begin by motivating our model with RG arguments and derivations of
it in two limits.    The rest of the paper explains how we  
{\it calculated}   
the phase diagram shown in Figure \ref{Figure1}.

\begin{figure}[htb] 
\begin{center}
\hspace{-15mm}
\psfrag{A}{$k_x$ }
\psfrag{B}{$k_y$}
\psfrag{p}{$\pvec$}
\psfrag{k}{$\kvec$} 
\psfrag{f}{$\kvec_F$}
\psfrag{P}{$\pi$}
\includegraphics[width=5cm]{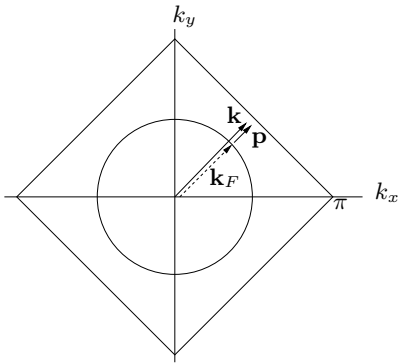} 
\end{center}
\caption{Expansion around a circular Fermi Surface.  The diamond
corresponds to a half-filled lattice.} 
\vspace{-2mm}
\label{Figure2} 
\end{figure}

Consider free non-relativistic particles with energy $\vep (\kvec)$
that is rotationally invariant,  e.g. $\vep(\kvec) = \kvec^2 / 2m_*$.
At finite density the Fermi surface is a circle in $2d$ as shown
in Figure \ref{Figure2}.  For a small band of energies near the Fermi surface,
$\kvec = \kvec_F (\kvec) + \pvec (\kvec)$ as shown
and the particles have a linear dispersion relation 
$\vep (\kvec ) \approx  \vep_F \pm v_F |\pvec|$.   Let $a_\pvec$
correspond to particles and $b_\pvec$ to holes.  Then the effective
hamiltonian is 
\beq
\label{E.8}
H =  \int_{|\pvec| < \pcut}  (d^2 \pvec)  \[ 
(v_F |\pvec | - \muhat )  \adag_\pvec a_\pvec + 
(v_F |\pvec| + \muhat ) \bdag_\pvec b_\pvec  \]  
\eeq
where
$\muhat = \mu - \vep_F$ is zero at zero temperature.

Our expansion around the Fermi surface is in the same spirit
as in\cite{FermiSurface},  but we now depart from it by 
demanding  a consistent local effective quantum field theory
that reproduces the above $H$.    Since the energy corresponds to massless
particles with a linear dispersion relation,  we identify
an emergent Lorentz symmetry and describe $H$ using
a relativistic field theory.  There are only 2 known candidates which 
differ in whether the lagrangian is first or second order in 
derivatives.   The first order case requires a multi-component
Dirac field, and since here a 4-fermion interaction is an irrelevant 
 dimension $4$ operator
in $2d$,  it cannot lead to a non-Fermi liquid.  Furthermore, 
Dirac fermions usually require additional properties of the Fermi surface
(Dirac points) 
as in graphene, where the multi-components have to do with 
sub-lattices.   

The other possibility is second-order in both space and time derivatives
with action
\beq
\label{II.1}
S = \int  dt \, d^2 \xvec  \(  \d_t \chi^- \d_t \chi^+ - v_F^2 
\gradvec \chi^- \cdot \gradvec \chi^+  \) 
\eeq
The Fermi velocity plays the role of the speed of light which
just serves to convert units of space and time so it can be set equal
to 1.  
This  form for a fermionic field is very unconventional,
and to a particle physicist it appears to violate the spin-statistics
theorem.    Since the above kinetic
term is crucial to all that follows,  we give the following 
compelling arguments in favor of it:
(i)  It correctly reproduces the desired effective hamiltonian 
(\ref{E.8}) for particles and holes near the Fermi surface.  
(See below.)   Thus the free theory is perfectly hermitian and
unitary in momentum space,  i.e. has no negative norm states.  
(ii)  In the condensed matter context, spin is a flavor
and for spin $\inv{2}$ particles we simply double the number of
components $\chi^\pm_\up , \chi^\pm_\down$.   Since there is a total
of 4 fields,  by Fermi statistics there is a unique 4-fermion interaction
with hamiltonian density 
\beq
\label{II.2}
\CH_{\rm int}  =   8 \pi^2 g  
 ~ \chi^-_\up \chi^+_\up \chi^-_\down \chi^+_\down
\eeq
Repulsive interactions correspond to positive $g$.  
Since the field has classical scaling dimension $\inv{2}$ in $2d$,  the
above operator has dimension $2$ and is therefore relevant.    
At low energies the coupling flows to an interacting fixed point 
with non-Fermi liquid behavior.   
 (iii)    Although our model was originally motivated by expanding
near a circular Fermi surface,  it can also be obtained from 
interacting itinerant lattice fermion models like the Hubbard
model at half-filling,  thus it can interpolate between the two Fermi
surfaces shown in Figure \ref{Figure2}.  At strong coupling the latter is known
to correspond to the Heisenberg anti-ferromagnet, with a low
energy description in terms of an $O(3)$ non-linear sigma model 
for a field $\phivec$ constrained to have fixed length with  lagrangian 
$\CL =   \d_\mu \phivec \cdot \d^\mu \phivec $\cite{Haldane,Affleck}. 
  In our model the $\phivec$ order parameter
is $\phivec = \chi^- \sigmavec \chi^+ /\sqrt{2}$ (see below)
and the constraint on $\phivec$ follows from the  simple constraint
 $  \chi^- \chi^+$ equal to a constant.  
  Inserting this into the $\phivec$  action one obtains
the second order action  (\ref{II.1}) up to some 
irrelevant operators.  This was pointed out in \cite{LeClair1}
and explained in more detail below  in connection with doping. 
There is no similar construction for Dirac fermions.

There is no violation of the spin-statistics connection since spin
is a flavor in this context and the Pauli-exclusion principle is
built in from the fermionic nature of the $\chi$-fields.   
The issue rather has to do with unitarity.   The mode expansion of
the fields is 
\barray
\nonumber
\chi^- (\xvec ,t )  &=& \int \frac{ (d^2 \pvec)}{ \sqrt{2\omega_\pvec}}
\(  \adag_\pvec  \, e^{-i p \cdot x }   +  b_\pvec \,  e^{i p \cdot x}  \) 
\\
\label{S.15}
\chi^+ (\xvec ,t )  &=& \int \frac{ (d^2 \pvec)}{ \sqrt{2\omega_\pvec}}
\( - \bdag_\pvec  \, e^{-i p \cdot x }   +  a_\pvec \,  e^{i p \cdot x}  \) 
\earray
where $\omega_\pvec = \sqrt{\pvec^2}$ and 
$p\cdot x \equiv \omega_\pvec t - \pvec \cdot \xvec $.   
The additional minus sign in the expansion of $\chi^+$ is chosen
so that the canonical quantization relations of the fields
leads to the usual  canonical relations in
momentum space for the $a$ and $b$'s. 
The canonical hamiltonian of the theory is precisely
(\ref{E.8}).  
Introduce a unitary operator $C$ that distinguishes particles and 
holes:  $C a C = a$, $ C b C = - b$ where $C^2 =1$.   
Then $\chi^+ = C (\chi^- )^\dagger C$ and in terms of fields
the hamiltonian is pseudo-hermitian: $H^\dagger = C HC$\cite{Neubert}.  
It was understood long ago by Pauli  that 
a pseudo-hermitian hamiltonian gives a consistent quantum mechanics
with a  unitary time evolution and real eigenvalues.  
In the present context,  pseudo-hermiticity has additional meaning
with regard to the kinematics of the expansion around the Fermi surface
since $C$ distinguishes particles and holes.   Conservation of
the physical momentum $\kvec$ is only equivalent to conservation 
of $\pvec$ for processes where particles are paired with particles 
and holes with holes.   For the study of SC,  these are of course
the processes we are primarily concerned with.   We are thus
only interested in eigenstates which are also eigenstates of $C$,
and for these $H = H^\dagger$.

The $SO(5)$ symmetry  is easiest to see
if one introduces an $N$-component version with fields
$\chi^\pm_\alpha$, $\alpha = 1,..,N$, which 
has $Sp(2N)$ symmetry\cite{LeClair1}.  For spin $\inv{2}$ particles,
$N=2$ and  $Sp(4) = SO(5)$.   
The $SO(5)$ has an $SU(2)$ spin subgroup and a $U(1)$ charge that
commutes with it.   The $\pm$ indices on the fields
$\chi^\pm$ correspond to electric charge. 
One can construct an $SO(5)$ vector of order parameters
$\vec{\Phi} = (\phi_x , \phi_y , \phi_z , \phi^+_e , \phi^-_e )$
where $\phivec$  
is an electrically neutral $SU(2)$ vector 
and $\phi^\pm_e$ are Cooper pair fields of charge $\pm 2$ 
which are $SU(2)$ spin singlets:
\beq
\label{orderparams}
\phivec =  \chi^- \sigmavec \chi^+ /\sqrt{2}, ~~~~~
\phi^+_e  = \chi^+_\up \chi^+_\down ,  ~~~~~~
\phi^-_e  = \chi^-_\down \chi^-_\up 
\eeq
The interaction can be written in a manifestly $SO(5)$ invariant
fashion:
$\CH_{\rm int} = -\frac{8 \pi^2 g}{5} ~ \vec{\Phi} \cdot \vec{\Phi}$.

It is important to carry out the RG directly in $2d$.   
As usual,  the RG prescription involves two energy scales,
the cut-off $\Lambdac$ and a lower running scale $\Lambda$. 
In many calculations $\Lambda$ is a lower cut-off. 
Since the coupling $g$ has units of energy in $2d$, we define
$g(\Lambda) = \Lambda \ghat (\Lambda)$ where $\ghat$ is dimensionless.
The 1-loop beta function is 
$- \frac{ d \ghat}{\log \Lambda}  =  \ghat - 8 \ghat^2$ 
  which has a low energy fixed point at $\ghat_* = 1/8$. 
To understand the phase diagram as a function of doping,  
it is first convenient to introduce the variable $x=1/\ghat$
where the fixed point value is at $x_* = 8$.    
We also introduce a variable $x_0$ that encodes the strength of
the coupling at the cut-off:  $x_0 = 1/\ghat_0 $ where 
$g(\Lambdac) = \Lambda_c \ghat_0$.    We assume that
at short distances  the coupling is strong,  i.e. $g > g_*$.   
It will also be useful  to define $\gamma = (x_* - x_0)/x_*$
which is a small parameter between $0$ and $1$   
The coupling at short distances can be arbitrarily strong,
where infinite coupling corresponds to $\gamma =1$.   
Integrating the beta-function with this  initial short-distance
data gives a linear form that is specific to $2d$
\beq
\label{scale}
T_{pg} \equiv 
 \frac{\Lambda}{\Lambda_c} = -\inv{\ratio} \( \frac{x}{x_*} - 1 \) 
\eeq
and  turns  out to be important in connection with hole  doping,
which we now turn to.

In the non-linear sigma-model description at half-filling,
the order parameter $\phivec$ is constrained to have fixed length.
As explained above  this constraint follows from
a constraint on the $\chi$ fields:
$  \chi^-_\up  \chi^+_\up + \chi^-_\down \chi^+_\down   
= i h \Lambda_c $.    Then one can show 
$\phivec \cdot \phivec = 3 h^2 \Lambda_c^2 /2$.    
Relaxing this constraint moves away from half-filling.   Thus
a measure of hole doping is the 1-point function
$h = -i  \langle \chi^- \chi^+ \rangle / \Lambda_c$.
The overall scale of $h$ as a doping variable can be justified by
reintroducing the chemical potential and noting that near the Fermi
surface it couples to the above operator.   
Including the 1-loop  order $g$  self-energy correction  to the propagator 
and expressing everything in terms of the $x$ variables one
obtains
\beq
\label{hofx}
h(x) = \inv{\pi^2} \( \frac{x-x_0}{x_* - x_0 } \) \[ 1 + \frac{4}{x}
 \( \frac{x-x_0}{x_* - x_0 }\) \] 
\eeq
In Figure \ref{Figure1}  the straight line is a plot of $\Lambda / \Lambda_c$ 
in eq. (\ref{scale})  as a function of the above $h$ for 
$x_0 =0$,  i.e. $\gamma =1$. (For $\gamma \neq 1$ it
is not exactly a straight line.)   It crosses the $h$ axis 
at $h_* = 3/2 \pi^2$,  which is the location of the RG fixed point.  
Below this line,  the energy scale is such that 
electrons described by our $\chi$ fields are strongly correlated
and exhibit non-Fermi liquid behavior since the fixed point is
not a free field theory.   Only $\langle \chi^- \chi^+\rangle \neq 0$
and no symmetries are broken at this line.
The  region below is what is normally called the pseudogap.   
The scale $\Lambda$ can be associated with a pseudogap temperature $T_{pg}$.

The AF phase can be analyzed by a standard mean field analysis.
One introduces an auxiliary field $\svec$ for the order parameter
$\phivec$ and derives the effective potential for constant
$\svec$  by performing the functional integral over the $\chi$ fields.
Minimizing this effective potential with respect to $\svec$ 
gives the gap equation
\beq
\label{AFgap}
\svec  =  - 16 \pi^2 g   \int_0^{\Lambda_c}  \frac{d\omega 
\, d^2 \kvec }{(2\pi)^{3}}
\, \frac{\svec}{ (\omega^2 + \kvec^2 )^2  -\svec^2  }
\eeq
For positive $g$, there are solutions due to the compensating
minus signs.   Since $s$ has dimension $2$, define
$s = \deltas^2 \Lambda_c^2$.  Then $\deltas$ is a solution to
the equation  
\beq
\label{AFgapdeltas}
\frac{\Lambda_c}{g} = \frac{4}{\deltas} \( \inv{2}  
\log \( \frac{\deltas + 1}{\deltas -1} \) - \tan^{-1} 1/\deltas \)
\eeq
When $g$ is small enough,  the solution flattens out with 
$\deltas \approx 1^+$.   This behavior is unphysical since
the gap should be zero when $g$ is zero.  The resolution of this puzzle
involves regulating the infra-red divergence with the low energy
cut-off $\Lambda$ and interpreting the result with the RG.   
Setting $s=0$, the gap equation can be approximately re-expressed
as $1/g(\Lambda) = 8 / \Lambda_c$.    This shows that at $g = \Lambda_c /8$,
a consistent non-trivial 
solution is $s=0$.   We interpret this as a first-order
transition where $\deltas$ drops discontinuously to zero.    
In terms of $x$ this occurs at
$x_{AF}  =  \frac{x_*}{1+\gamma}$.
Since $\deltas$ is in units of the cut-off it is meaningful
to rescale it and define $\deltas' = \frac{\Lambda}{\Lambda_c} \deltas$,
where the scale factor is given in terms of $x$ in (\ref{scale}).   
In Figure \ref{Figure1}  we show the solutions to the gap equation as a function
of doping $h$.  The N\'eel temperature $T_N$ is proportional to the
zero temperature gap as we will describe below.   The AF 
transition occurs at $h_{AF} = h(x_{AF}) = 3/4\pi^2$ when $\gamma=1$.

Introducing constant auxiliary fields $q^\pm$ for the SC order parameters
$\phi^\pm_e$ and repeating the above mean field analysis 
gives a gap equation for $q^2 = q^+ q^-$ that has the same form 
as (\ref{AFgap}) with $s^2 \to -q^2 $.  This leads to the expected
result: for repulsive interactions (positive $g$) there are no solutions,
i.e. no s-wave SC.   This shows that the AF and SC phases do not compete,
and the SC phase is not simply related to the AF one by the $SO(5)$ 
symmetry.   
When one goes beyond mean field 
and  incorporates momentum dependent scattering of Cooper pairs
near the Fermi surface, an attractive d-wave channel opens up.  
Introducing  non-constant auxiliary pair fields, 
 one can derive the momentum dependent gap equation 
\beq
\label{dwaveq}
q(\kvec) =  -  \int \frac{ d\omega \, d^2 \kvec'}{(2\pi)^{3}} 
~  G(\kvec , \kvec' ) ~  \frac{ q(\kvec' )}{(\omega^2 + \kvec'^2)^2 + 
q(\kvec' )^2 }
\eeq
where $q^+ = q^- $ up to a phase.  
The kernel $G$ is related to a particular 4-particle Green function 
specialized to Cooper pairs of opposite 
momenta $\pm \kvec$ and $\pm \kvec'$.     
In a rotationally invariant theory one can expand in  circular harmonics:
\barray
\nonumber
G(\kvec , \kvec') &=& \sum_{\ell =0}^\infty G_\ell (k,  k') ~
\cos \ell (\theta - \theta')
\\ \label{harmonics}
q (\kvec ) &=& \sum_{\ell =0}^\infty  q_\ell (k) \cos \ell \theta
\earray
where $k$ is the magnitude of $\kvec$ and $\theta - \theta'$ is
the angle between $\kvec $ and $\kvec'$.   
Performing a low energy momentum expansion at 1-loop one finds an attractive $\ell =2$ channel:
$G_2 (k , k') = - 8 \pi^2 g_2  k^2 k'^2 $
where $g_2 = 4 \ghat^2 / 25 \Lambda^3$.   
The solution to the gap equation has the characteristic d-wave form
\beq
\label{dwaveform}
q (\kvec ) =  \deltaq^2  k^2  \cos 2\theta =  \delta_q^2 (k_x^2 - k_y^2) 
\eeq
where $\deltaq$ is a constant solution to the 
integral equation 
\beq
\label{orb.7}
\deltaq^4 =  2 g_2 \int_0^\pcut   d\omega \,  dk^2 \,  
\(  1 -  \frac{ \omega^2 + k^2 }
{\sqrt{ (\omega^2 + k^2 )^2 + \deltaq^4 k^4}}  
\)
\eeq
To plot the gap as a function of doping one needs to express the  constant $g_2$  in terms of $x$:
$g_2 =  \inv{\Lambda_c^3}  \frac{4 \gamma^3}{25 x^2 (1-x/x_*)^3}$,
and one sees that it changes sign at $x=x_*$,  which implies that
the SC phase terminates at the RG fixed point.   
The other important feature of the above gap equation for $\deltaq$ 
is that SC turns off at a lower, non-universal value of
 $x_1 \approx ( 1 - 0.1\gamma) x_*$.   
Thus the non-zero solutions  of the d-wave  gap equation fall under a dome
$x_1 < x < x_*$.   Comparison with experiments shows that
this dome as calculated is too narrow;  however there are many
effects we have neglected that could broaden it, e.g. interplane coupling,
disorder, or higher order corrections.  
Numerical study of solutions of (\ref{orb.7}) indicates that 
optimal doping occurs in the tight range between $0.13 $ and $0.15$ 
where the upper bound is simply $h_*$.   
Solutions to the d-wave gap equation
$\deltaq' = \Lambda \deltaq / \Lambda_c$  are shown in Figure \ref{Figure1}.

As in the BCS theory, one expects that the critical temperature $T_c$ 
is proportional to the zero temperature gap.    Rather than
developing a full-blown finite temperature formalism, one
can study the effect of a small finite temperature by introducing
a relativistic mass term in the lagrangian $-m^2  \chi^- \chi^+$
where $m = \alpha T$ with $\alpha$ a dimensionless constant.   
The justification for this is that in the free theory this thermal
perturbation reproduces the correct electronic specific heat 
$C_V \propto T k_F / v_F$.   Requiring the overall coefficient to 
match the standard non-interacting  result extended to  $2d$ fixes
$\alpha = \pi^{5/4} / \sqrt{6} \approx 1.7$.   
We have verified numerically that the solutions of the AF and d-wave
gap equations with this additional mass term, $\omega^2 \to \omega^2 + m^2$,
vanish when $m$ is too large,  and the critical value
$m_c = \alpha T_c$ is indeed proportional to the zero temperature gap
up to constants of order 1 that depend weakly on $\gamma$.  
Restoring fundamental constants and the Fermi velocity $v_F$ one
then finds
$k_B T_c  =  \frac{c}{\alpha}  ~  v_F \hbar \deltaq' \Lambda_c$.
At optimal doping the constant $c$ is in the range $.5 < c < .7$
depending on $\gamma$ and $\deltaq' \approx 0.11$ is
largely  insensitive to $\gamma$. 
A similar formula applies to the N\'eel temperature $T_N$ 
with $c\approx 1$.  
The Fermi velocity $v_F$ in our relativistic model 
is a fixed constant
that plays the r\^ole of the speed of light which is universal
in the sense that it is independent of the coupling and thus the doping $h$. 
Remarkably, at low energies 
a doping independent universal nodal Fermi velocity 
has been observed in recent years\cite{FermiV}.   This can actually
be viewed as a prediction of our theory, where the nodal direction
$\kvec$ from $(0,0)$ to $(\pi , \pi )$  shown in Figure \ref{Figure2} is what is important since
this is the direction where the half-filling diamond is closest 
to the nearly circular Fermi surface just inside it.  
Using the results in\cite{FermiV}  we 
estimate $v_F \approx 1.4 ev \AA = 210 km/s$ for LaSrCuO.  
This leads to an estimate of $T_c$ at optimal doping.
The cut-off $\Lambda_c$ should be equated with the inverse lattice
spacing $a$.   A practical form  is
\beq
\label{Tcev}
T_c  =  c\, \frac{v_F}{a} \cdot 650K
\eeq
where $v_F$ is in $ev \AA$ and the lattice spacing in $\AA$
and we have set $\alpha$ to its estimated value of $1.7$.   
With the above $v_F$ and $a=3.8 \AA$ for the CuO square lattice,
this gives $T_c $     in the range  
$120K < T_c < 160K$ depending on $\gamma$.   
The maximum $T_c$ occurs around $\gamma = 1/4$. 
The simple result (\ref{Tcev})  provides some insights on how 
to possibly obtain a higher $T_c$:  shorten the lattice spacing 
or increase the Fermi velocity by possibly reducing the effective
electron mass $m_*$.   

Many more properties can be calculated,  such as anomalous corrections
to the specific heat,   a resistivity linear in temperature,  and electron
correlation functions in the pseudgap region.   These additional results
and  a more detailed exposition  will be be presented in the 
comprehensive article\cite{dwave2}.

We would like to thank Seamus Davis,  Kyle Shen and Henry Tye for
discussions.    This work is supported by the National Science Foundation.


\begin{thebibliography}{99}



\bibitem{Anderson}   P. W. Anderson,  
Science {\bf 256} (1992)  1526.


\bibitem{Zhang}   S. C. Zhang,  Science {\bf 275} (1997) 1089.  




\bibitem{LeClair1}   A. LeClair, 
arXiv:cond-mat/0610639,  arXiv:cond-mat/0610816.



\bibitem{Neubert}  A. LeClair and M. Neubert, 
JHEP {\bf 10} (2007) 027.



\bibitem{Anderson2}  G. Baskaran, Z. Zou and P. W. Anderson, Sol. State. Commun. 
{\bf 63}  (1987) 973.

\bibitem{YangZhang}  C. N. Yang and S. C. Zhang,  Mod. Phys. Lett. {\bf B4}  (1990) 
759. 


\bibitem{Vojta}  M. Vojta and S. Sachdev,  Phys. Rev. Lett. {\bf 83} (1999) 3916. 


\bibitem{FermiSurface}   G. Benfatto and G. Gallavotti,  J. Stat. Phys. 
{\bf 59} (1990) 54; 
  R. Shankar,   
Rev. Mod. Phys. {\bf 66} (1994) 129; 
 J. Polchinski,   [hep-th/9210046] ; 
S. Weinberg, 
Nucl. Phys. {\bf B413}  (1994) 567. 












\bibitem{Haldane}  F. D. M. Haldane,  Phys. Lett. {\bf 93A} (1983) 464;
Phys. Rev. Lett. {\bf 50} (1983) 1153. 



\bibitem{Affleck}   I. Affleck, Les Houches lectures 1988,
North-Holland.   

\bibitem{FermiV}  X. J. Zhou et. al. Nature {\bf 423} (2003) 398.


\bibitem{dwave2}   E. Kapit and A. LeClair,  {\it A model of a $2d$ non-Fermi
liquid with $SO(5)$ symmetry,  AF order,  and a d-wave SC gap},   to appear. 


\end{thebibliography}
\end{document}